# Thermal stability and anisotropic sublimation of two-dimensional colloidal $Bi_2Te_3$ and $Bi_2Se_3$ nanocrystals

*Joka Buha[†,\*], Roberto Gaspari[‡], Antonio Esau Del Rio Castillo[⊥], Francesco Bonaccorso[⊥], Liberato Manna[†,\*]*

[†]Department of Nanochemistry, Istituto Italiano di Tecnologia, Via Morego 30, 16163 Genova, Italy

[‡]CompuNet, Istituto Italiano di Tecnologia, Via Morego 30, 16163 Genova, Italy

[⊥]Graphene Labs, Istituto Italiano di Tecnologia, Via Morego 30, 16163 Genova, Italy

ABSTRACT

The structural and compositional stabilities of two-dimensional (2D) $Bi_2Te_3$ and $Bi_2Se_3$ nanocrystals, produced by both colloidal synthesis and by liquid phase exfoliation, were studied by *in situ* transmission electron microscopy (TEM) during annealing at temperatures between 350°C and 500°C. The sublimation process induced by annealing is structurally and chemically anisotropic and takes place through the preferential dismantling of the prismatic $\{01\bar{1}0\}$ type planes, and through the preferential sublimation of Te (or Se). The observed anisotropic sublimation is independent of the method of nanocrystal's synthesis, their morphology, or the presence of surfactant molecules on the nanocrystals surface. A thickness-dependent depression in the sublimation point has been observed with nanocrystals thinner than about 15 nm. The $Bi_2Se_3$ nanocrystals were found to sublimate below 280°C, while the $Bi_2Te_3$ ones sublimated at temperatures between 350°C and 450°C, depending on their thickness, under the vacuum conditions in the TEM column. Density Functional Theory calculations confirm that the sublimation of the prismatic $\{01\bar{1}0\}$ facets is more energetically favourable. Within the level of modelling employed, the sublimation occurs at a rate about 700 times faster than the sublimation of the $\{0001\}$ planes at the annealing temperatures used in this work. This supports the distinctly anisotropic mechanisms of both sublimation and growth of $Bi_2Te_3$ and $Bi_2Se_3$ nanocrystals, known to preferentially adopt a 2D morphology. The anisotropic sublimation behaviour is in agreement with the intrinsic anisotropy in the surface free energy brought about by the crystal structure of $Bi_2Te_3$ or $Bi_2Se_3$.

Bismuth chalcogenides, $Bi_2Te_3$ and its alloys in particular, have been amongst the best established thermoelectric (TE) materials for decades.[1,2] In recent years, both $Bi_2Te_3$ and $Bi_2Se_3$ have attracted a renewed surge of interests as topological insulators,[3] opening doors for many possible new optical and electronic applications, such as spintronics and superconductor-topological insulator devices for quantum computing.[4,5] The nanostructured Bi-chalcogenides are particularly advantageous over the bulk counterparts for the TE applications because a reduction in crystal dimensionality allows their electronic and thermal properties to be tuned independently, while a large number of crystal interfaces may additionally effectively scatter phonons further reducing the thermal conductivity.[6-8] Both $Bi_2Te_3$ and $Bi_2Se_3$ exhibit a layered structure consisting of quintuple layer (QL) segments separated by a van der Waals (vdW) gap with each QL consisting of five consecutive Se or Te and Bi layers (Figure 1a).[9,10] In addition to methods based on physical vapour deposition[11] vapour-liquid-solid deposition,[12] vapour-solid deposition,[13] molecular beam epitaxy,[14] mechanical exfoliation[15] and liquid-phase exfoliation (LPE)[16-18], several solution-based processes for the synthesis of ultrathin colloidal Bi-chalcogenide nanocrystals (NCs) have also been developed.[19-24] The technological application of such colloidal NCs is however still at an early development stage.[25]

For the technological applications, the TE ones in particular, the Bi chalcogenide-based devices have to be structurally and chemically stable at service temperatures, currently limited to less than about 350°C.[2] The properties of topological insulators are also known to be dependent on the presence of structural defects and variations in the chemical composition,[26] both generally sensitive to thermal treatment. The integration of colloidal NCs into devices generally involves thermal treatments such as sintering, ligand removal, annealing and deposition of metal contacts.[25,27,28] Although the structural and chemical integrity of $Bi_2Te_3$ and $Bi_2Se_3$ NCs at elevated temperatures is critical for technological applications, it has not yet received the due attention. Even the most fundamental thermally induced phase transitions in $Bi_2Te_3$ and $Bi_2Se_3$ NCs, such as sublimation or melting, have not been observed before on an atomic scale, so the mechanisms involved are not clear. The observations of crystal melting or sublimation are additionally insightful as they may reveal processes and mechanisms that could be acting also during the reverse process, i.e. that of crystal growth. To date, studies on melting and sublimation of anisotropic colloidal NCs have been limited to one-dimensional (1D) NCs [29-33] and branched nanostructures (CdSe/CdS octapods).[34] Additionally, recent studies on the mechanically exfoliated phosphorene[35] and graphene[36,37] in particular, reveal that annealing gives rise to unique structural changes, such as multiple-layer edge reconstructions, not found in the three-dimensional (3D) bulk systems.[36,37] Overall, there is therefore a need to fully understand the thermal stability of 2D colloidal NCs from both practical and fundamental points of view. Here, we report on the thermal stability and sublimation mechanisms of the 2D $Bi_2Te_3$ and $Bi_2Se_3$ NCs, which were studied in solid state by means of *in situ* transmission electron microscopy (TEM). The study, including the NCs produced by colloidal syntheses and by the LPE, indicates that the 2D nanocystals exhibit reduced thermal stability as compared to bulk counterparts. Our findings also shed light on the relative stabilities of different crystallographic facets and, based on the observed anisotropic sublimation mechanism and the supporting density functional theory (DFT) calculations, also on the growth mechanism of the $Bi_2Te_3$ and $Bi_2Se_3$ nanocrystals in solid state systems.

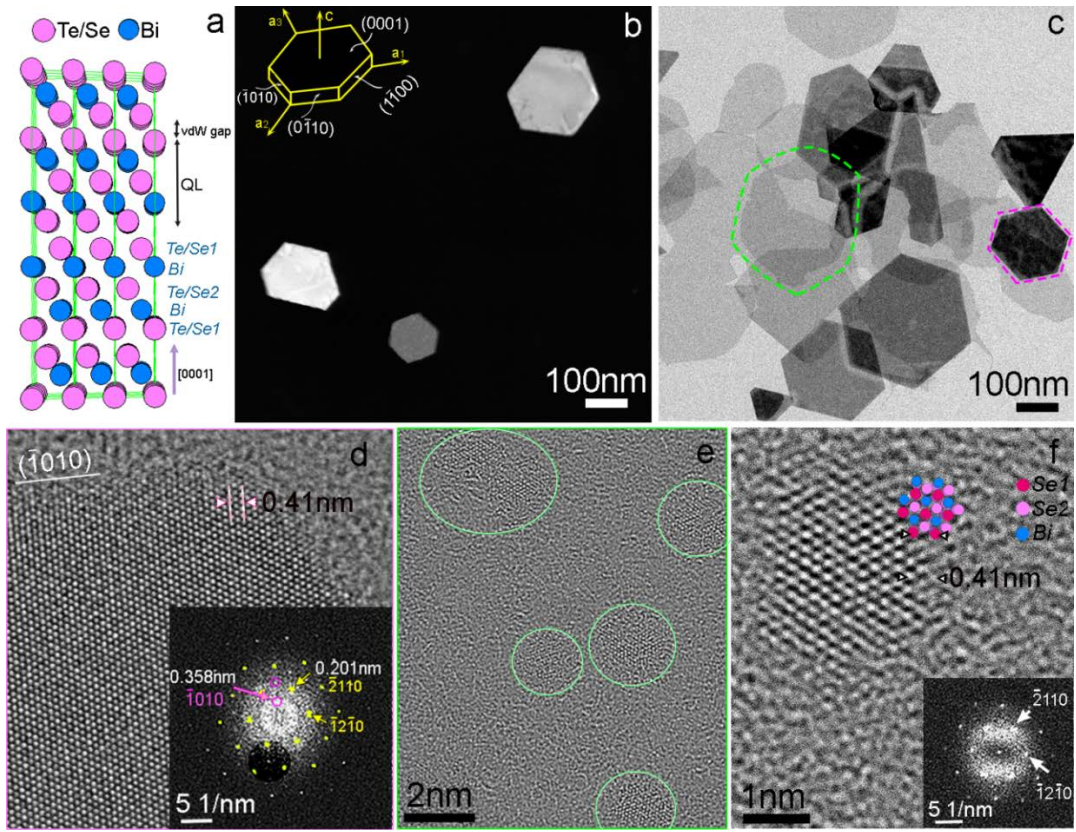

***Figure 1.*** *The as-synthesized colloidal $Bi_2Te_3$ and $Bi_2Se_3$ NCs. a) The unit cell of the rhombohedral crystal structure of $Bi_2Te_3$ and $Bi_2Se_3$. b) STEM image the $Bi_2Te_3$ NCs; their crystallographic orientation with respect to substrate is sketched in the inset; c) TEM image of the $Bi_2Se_3$ platelet-like NCs (darker contrast and hexagonal morphology; an example is outlined by the purple broken line) and thin nanosheets (faint contrast and poorly defined morphology in lateral directions; an example is outlined by the green broken line); d) HRTEM image from one of the $Bi_2Se_3$ platelet-like NC in its [0001] zone axis with the corresponding Fourier transform (FT) in inset; e) HRTEM image from one of the amorphous $Bi_xSe_y$ nanosheets with isolated crystalline domains within (circled). One of these domains is shown enlarged in f), along with the corresponding FT and a sketch of the arrangement of Bi and Se atoms as visible in the [0001] zone axis (insets). Note that the forbidden $01\bar{1}0$ reflections may appear in the electron*
 

*diffraction patterns and FTs due to multiple scattering of the electron beam passing through the NC.*

The colloidal $Bi_2Te_3$ and $Bi_2Se_3$ platelet-like NCs synthetized in the presence of poly(vynilpirrolidone) (PVP) according to an established procedure[23] were primarily used for the study. The PVP molecules form a surfactant layer on the NCs which is expected to be stable up to about 350°C when its decomposition commences,[38,39] leaving behind a carbon shell.[40] In order to eliminate any surface-stabilizing effect of the PVP, three additional $Bi_2Te_3$ samples were included in the study: (i) the colloidal NCs synthetized in the presence of PVP from which the PVP shell was subsequently removed by means of a hydrazine-based washing procedure;[24,41] (ii) the colloidal NCs synthetized following the same procedure[23] but without the presence of any surfactant; and (iii) $Bi_2Te_3$ flakes prepared by the LPE of bulk $Bi_2Te_3$ in a mixture of isopropanol and water.[42,43] The *in situ* thermal treatment experiments were performed in the column of an aberration corrected JEOL JEM 2200FS microscope operated at 200 kV and under the vacuum of $1.8 \times 10^{-5}$ Pa. The NCs were characterized by high resolution TEM (HRTEM), scanning TEM (STEM) and energy-dispersive X-ray (EDX) spectroscopy. The *in situ* heating in the TEM column was carried out at temperatures between 350°C and 500°C for the periods of time between 30 to 120 minutes, as indicated in the text. More details on the annealing experiments and TEM characterization, as well as the additional characterization of all samples by means of X-ray diffraction (XRD), Raman spectroscopy and optical absorption spectroscopy are presented in the supporting information (SI). The kink energies of the relevant crystal surfaces were computed by DFT[44,45] calculations on periodic slab geometries of $(0001)$ and $(01\bar{1}0)$ terminated $Bi_2Te_3$ models.

**The structure of the colloidal $Bi_2Se_3$ and of $Bi_2Te_3$ NCs.** The colloidal synthesis yielded platelet-like $Bi_2Te_3$ and $Bi_2Se_3$ NCs of mostly hexagonal shape, with their basal {0001} planes parallel to the nanoplatelet face (Figure 1b) and 1c)) and their edges terminated by the prismatic {01$\bar{1}$0} type planes (inset in Figure 1b)), which indicates that the growth in the ⟨01$\bar{1}$0⟩ directions is favoured and occurs faster than the growth along other low-index crystallographic directions. The thickness of most platelet-like NCs was up to approximately 15 nm (as determined from TEM observations of the NCs oriented edge-on). The well-defined thicker NCs (giving out darker contrast in TEM) and occasional thinner sheet-like NCs having larger lateral size, were obtained during the synthesis of both $Bi_2Te_3$ and $Bi_2Se_3$ (Figure 1b) and 1c)). In the case of $Bi_2Se_3$ however, the very thin nanosheets (one is outlined by a green broken line in Figure 1c) were mostly amorphous with isolated domains of crystalline phase of 2-3 nm in lateral size (Figure 1e) and 1f)). These exhibited the structure matching the $Bi_2Se_3$ phase in the lateral direction and were all oriented in their respective [0001] directions, albeit rotated a few degrees about the c axis with respect to each other (Figure 1f)). This indicates that the crystalline domains crystallize independently. The EDX analysis of such nanosheets (Figure S1) revealed that they contained both Se and Bi. The above observations suggest that the $Bi_2Se_3$ NC growth mechanism may be more diverse than proposed before[19] and may possibly involve also attachment of amorphous aggregates to growing nuclei. Structural defects such as the presence of the edge steps, incomplete layers and bending contours were observed in both $Bi_2Te_3$ and $Bi_2Se_3$ NCs.

**The thermal evolution of the PVP-capped $Bi_2Se_3$ and $Bi_2Te_3$ NCs.** Figure 2 shows the structural and compositional evolution of $Bi_2Te_3$ platelet-like NCs, synthesized in the presence of PVP, after a twofold annealing process, i.e., 30 min at 350°C followed by 30 min at 400°C.

Taking into account the relatively low melting points of bulk $Bi_2Te_3$ and $Bi_2Se_3$ (586°C and 710°C, respectively), the partial pressures of Te (~$10^{-1}$ Pa)[46], Se (> ~140 Pa)[46] and Bi (~ $10^{-6}$)[46] at the temperatures (~350°-500°C) and the vacuum ($10^{-5}$ Pa) used in this work, the *in situ* TEM heating of $Bi_2Te_3$ and $Bi_2Se_3$ NCs is expected to lead to their sublimation. Considering a basic temperature-pressure phase diagram, the sublimation temperature of a solid substance beyond the triple point on the negative pressure side (under vacuum) is expected to be lower than its melting temperature at the atmospheric pressure.

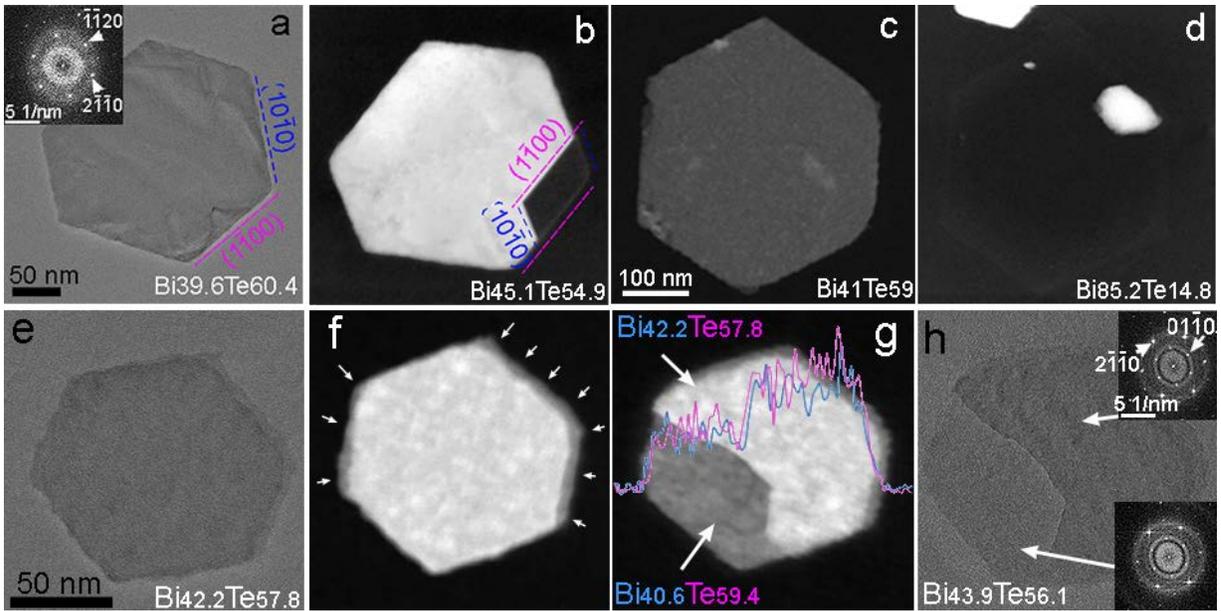

*Figure 2. The evolution of $Bi_2Te_3$ NCs during annealing at 350°C and 400°C. The chemical composition of the NCs determined by the EDX is given in at% of Bi and Te. a) TEM image of a platelet-like NC in the [0001] zone axis before annealing, with the corresponding FT in inset, and b) STEM image after a portion of the NC sublimated following annealing at 350°C. The sublimation front remained parallel to two of the $\{01\bar{1}0\}$ type planes indicated by broken lines. The compositional EDX analysis before and after annealing indicates the preferential sublimation of Te (a) and b); see also Figure S2). Image d) shows that in the final stages of*

*sublimation at 400°C leaving behind Bi-rich fragments, the sublimation front becomes rounded. Images e) to h) show an example of partial sublimation leading to a reduction in the thickness of the platelet-like NC, while the sublimation front remained parallel to $\{01\bar{1}0\}$ type planes. The STEM image in f) shows the thickness variations at the edges, most likely a defect formed during the NC growth. The EDX concentration profile of Bi and Te across the partly sublimated NC is superimposed on the image in g).*

During the annealing experiment on the $Bi_2Te_3$ NCs, as soon as the temperature approached 350°C several NCs underwent gradual sublimation starting from the edges and progressing along the $\langle\bar{2}110\rangle$ directions parallel to the $\{01\bar{1}0\}$ planes, consuming the entire thickness of the NCs (Figure 2b)). No noticeable changes in the crystal structure occurred as a result of the sublimation, however all NCs were Te-deficient as compared to their composition before annealing, which is an indication of the preferential Te sublimation. This means not only that sections of NCs sublimate entirely, but also that Te sublimates before Bi, and before the structural disintegration of NCs becomes obvious. This is consistent with the high $Te_{2(g)}$ vapour pressure above the molten $Bi_2Te_3$ phase.[47] The preferential Te sublimation may negatively impact on the TE applications,[2] considering also that the thermally-driven transition points are reduced in small NCs as compared to bulk crystal. The same phenomenon of preferential TE loss was observed during annealing between 200°C and 400°C with $Bi_2Te_3$ nanowires produced by pulsed electrodeposition,[48] and is known to occur with bulk material as well,[2] requiring encapsulation of the $Bi_2Te_3$ TE material.[49]

Figure 2e) to h) shows an example, where several layers along the c axis sublimated reducing the thickness of the platelet-like $Bi_2Te_3$ NC in that area. The sublimation in this case still involved the $\{01\bar{1}0\}$ planes. The difference in the composition between the thinner (dark) and thicker (brighter) regions of the NC after annealing at 350°C indicates that the sublimation involves individual atomic layers, rather than entire QLs. The mechanism of the $Bi_2Te_3$ NC sublimation is visible in more detail from the video S1, available in the SI, showing a corner of one of the platelet-like NCs sublimating at 400°C. The selected stills from this video, shown in Figure S4, indicate that the sublimation process takes place via gradual sublimation of individual atoms on the NC's edge along the $\langle\bar{2}110\rangle$ directions (visible as steps formed by $\{01\bar{1}0\}$ planes; indicated by a single purple arrow in the 1 min. 39 sec. and 1 min. 52 sec. frames in Figure S4a)), followed by the sublimation of the nearest neighbour atoms on the corresponding $\{01\bar{1}0\}$ type plane, as sketched in Figure S4b). This is illustrated in Figure 3, which shows different perspectives on a $Bi_2Te_3$ ($Bi_2Se_3$) NC at an advanced stage of sublimation; the differently coloured rows of atoms in Figure 3b) indicate the individual $\{01\bar{1}0\}$ planes.

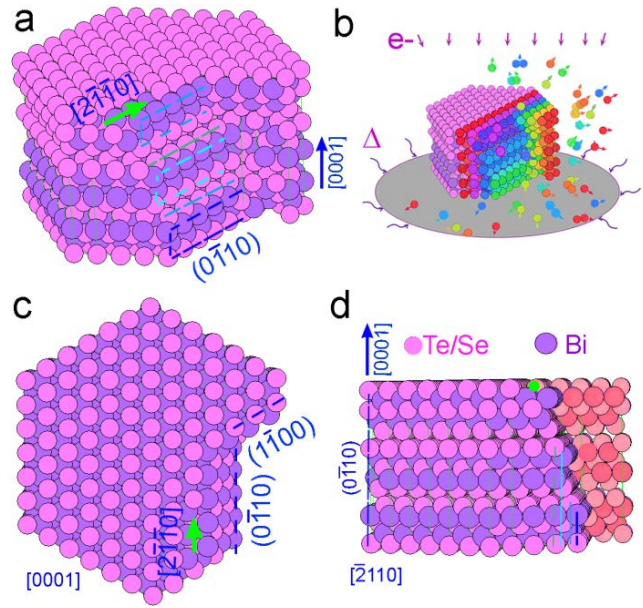

*Figure 3. The schematic representation of a unit cell thick NC following partial sublimation. a) The steps formed by the simultaneous sublimation of a number of $\{01\bar{1}0\}$ planes are indicated by broken lines; in b) the individual $\{01\bar{1}0\}$ planes undergoing sublimation are coloured*

*differently; c) top view (the [0001] zone axis) and d) side view (the [2̄110] zone axis) of the partially sublimated NC.*

An analogous preferential sublimation on the {011̄0} planes (the zigzag edges) was observed also during the annealing of few-layer graphene,[36,37] which inherits the hexagonal graphite arrangement of atoms and stacked C layers,[50] similar to that present in the $Bi_2Te_3$ ($Bi_2Se_3$) structure. The $Bi_2Se_3$ NCs followed a similar sublimation pattern during annealing, although the sublimation started at a lower temperature (below 280°C; Figure 4 a) to c)) as compared to the $Bi_2Te_3$ NCs. This could be linked with the higher vapour pressure of Se compared to Te and Bi,[46] leading to sublimation of Se at a lower temperature than Te under the vacuum conditions in TEM column. The thin and mostly amorphous nanosheets developed a pitting pattern during annealing with small areas from the interior and edges of the nanosheets sublimating (Figure 4 d) to f)).

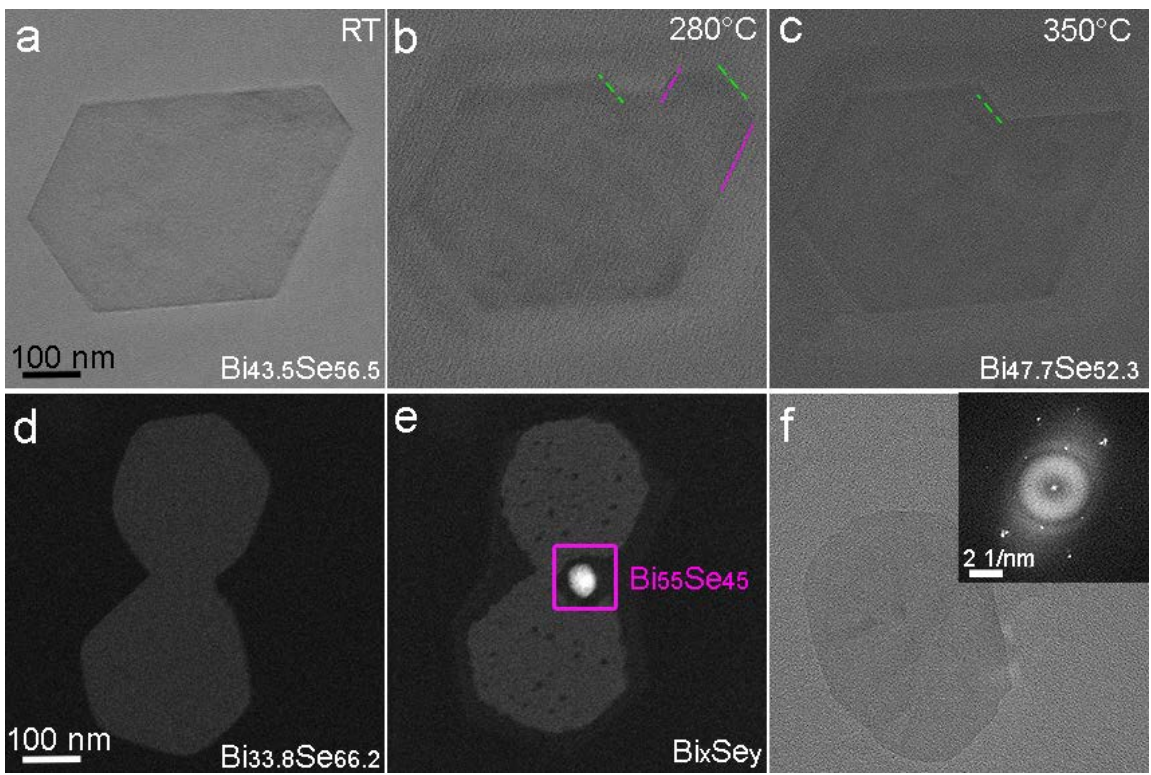

***Figure 4.*** *The evolution of the $Bi_2Se_3$ platelet-like NCs during annealing. a) to c) show the sublimation of a NC already underway at 280°C during the heating up, and progressing during annealing at 350°C. Panels d) to f) show the sublimation pattern developed on the thin nanosheets. The HRTEM image and the corresponding FT from the new phases outlined by a rectangle in e) and formed following the annealing at 350°C for 1 hour, are shown in f).*

These nanosheets remained amorphous after the annealing, which indicates that the nanocrystalline domains, observed within before annealing (Figure 1 e) and f)), sublimated preferentially. This observation highlights the surface instability related to anisotropic crystalline material as opposed to the amorphous surface of the thin nanosheets, which should exhibit a more isotropic surface free energy.[51] Some of the solute released from the nanosheets during the annealing precipitated nearby to form a new crystalline $Bi_xSe_y$ phase (Figure 4 e) and f); see also Figure S5 b) and c)).

Finally, it should be noted that the characterization of the annealed areas never exposed to the electron beam did not evidence significant differences from those annealed at the same temperature under illumination. This indicates that the electron beam irradiation during the *in situ* experiments (discussed in more detail in the SI) did not significantly affect the mechanism or the temperature of the sublimation of the NCs.

**The thermal evolution of the surfactant-free NCs.** The presence of the PVP is known to have an important role in stabilizing and controlling the morphology of metal-chalcogenide nanoparticles produced by different routes,[38] although the exact mechanisms have not been fully clarified. In addition to controlling the kinetics of different chemical reactions during the solution-based synthesis,[52] the PVP is also believed to preferentially bind to the outermost basal Te(Se) planes of $Bi_2Te_3$ ($Bi_2Se_3$) during their one-pot synthesis thus contributing to the platelet-like morphology of the crystals.[18,25] In order to eliminate any possible effect that the PVP may have on the stabilities of different crystal facets, the sublimation mechanism of three types of surfactant-free $Bi_2Te_3$ NCs was also investigated. Figure S6 shows that the sublimation mechanism of the colloidal $Bi_2Te_3$ NCs from which the PVP shell had been removed by the ligand-removal washing procedure[24,41] is analogous to that observed with the PVP-capped NCs. A similar behaviour was observed also with the platelet-like NCs produced by the surfactant-free synthesis (Figure S7). As considerably thicker NCs were obtained in this synthesis (approximately in the range of 15 to 90 nm; some are seen edge-on in Figure S7a)), no sublimation was observed at 350° or 400°C, while at 450°C and at 500°C the sublimation took place, but it was still at an early stage for most NCs. Compared with the thinner (up to about 15 nm thick) PVP-capped $Bi_2Te_3$ NCs (shown here in Figure 2), the observations made on the PVP-free NCs indicate that a size-dependent reduction in the phase transition point[53,54] applies at least for the thinner platelet-like NCs, observed here under the high vacuum of the TEM column. It is therefore expected that such thin platelet-like NCs would undergo thermally-driven phase transitions at lower temperatures compared to bulk material even at ambient pressure. Finally, even the randomly shaped $Bi_2Te_3$ flakes, produced by LPE from bulk material exhibited the same anisotropic sublimation behaviour involving the $\{01\bar{1}0\}$ planes (Figures S8, S9 and Video S2).

**The anisotropy of sublimation.** The experimental observations reported here show that sublimation of the 2D colloidal $Bi_2Te_3$ and $Bi_2Se_3$ NCs is distinctly anisotropic, taking place through the preferential sublimation of the $\{01\bar{1}0\}$ planes. The sublimation of the $\{01\bar{1}0\}$ planes takes place through the dissociation of individual atoms from the kink sites on those planes. The experiments furthermore indicate that the observed sublimation mechanism is independent of the presence of the PVP surfactants on the NCs surfaces, their morphology and the preparation method. The crystallographically anisotropic phase transitions involving melting and sublimation have been observed before[29,31,40,55-58] and theoretically predicted by the Lindemann's criterion.[69] This usually indicates anisotropy in the surface free energies of the different crystallographic planes. In the structure of $Bi_2Te_3$ ($Bi_2Se_3$), the $\{01\bar{1}0\}$ planes have the lowest packing density (the number of dimers per Å$^2$) hence the largest free surface energy compared with $\{0001\}$ and $\{\bar{2}110\}$ planes. As such, the $\{01\bar{1}0\}$ planes terminated facets are more likely to undergo sublimation or thermally induced disordering near the sublimation point before other facets. This is consistent with the principles of anisotropic surface melting of metals.[60,61] When it comes to colloidal NCs however, many other factors at play may potentially alter the order in which the different facets melt or sublimate. These factors could be the surface polarity, the presence of surfactant, the stability of the surfactant at the annealing temperature, the presence of contamination on the NC surface, the contact with substrate and/or another nanoparticle, surface relaxation, surface reconstruction, the nature of bonding in the crystal, etc., as discussed later in the text.

Since the present experiments show a consistent anisotropic sublimation mechanism acting in colloidal $Bi_2Te_3$ ($Bi_2Se_3$) NCs regardless of the presence of the surfactant, the preparation method or the starting morphology, DFT calculations were carried out to determine the energy barrier for the sublimation from the pristine (surfactant-free) $(0\bar{1}10)$ surface as compared to sublimation from the pristine $(0001)$ surface. The unit cells explicitly included in the calculations are shown in Figure 5. Since we aim here at rationalizing the rate-limiting step of the sublimation process, we only computed the kink energy of the most stable surface in the $\{0001\}$ family of planes, that is the surface obtained by cleaving the system at the van der Waals gap. This surface exposes Te atoms and the associated kink site was named Te$_{vdWG}$. On the other hand, the unit cell of the $(0\bar{1}10)$ surface exposes five inequivalent rows which are associated with kink sites named respectively Te$_A$, Te$_B$, Te$_C$, Bi$_A$, Bi$_B$, as reported in Figure 5. These five sites exhaust all possible inequivalent kink positions on the $\{01\bar{1}0\}$ type surfaces since the $\{01\bar{1}0\}$ layers only differ by a translation and display the same atomistic environment when exposed at the surface. The results of the calculations are summarized in Table 1. The kink energy of the Te$_{vdWG}$ position is the largest, amounting to 4.43 eV, followed by the Bi$_A$ site, with the energy of 4.06 eV. This indicates that, under the experimental conditions used in this work, the atomic desorption from the {0001} layers is remarkably slower at the Te layer containing the stable Te$_{vdWG}$ site, than the desorption along the atomic rows of the $\{01\bar{1}0\}$ surfaces.

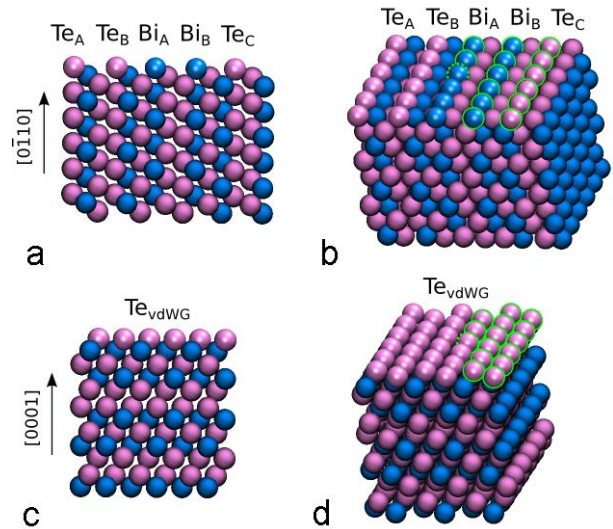

*Figure 5. The description of the* $(\mathbf{0001})$ *(a,b) and* $(\mathbf{0\bar{1}10})$ *(c,d) DFT slab models, shown in a lateral (a,c) and tilted (b,d) view. The Te atoms are light purple while Bi atoms are blue. Surface atoms are represented by reflecting spheres. The atoms bordered by the solid green circle are removed to create kink sites in the structure, the subfigures (b,d) representing the peculiar case*

*of the Bi$_A$ and Te$_{vdWG}$ kink sites. The sites bordered by the broken green circles represent the kink atoms, whose desorption energy is evaluated by DFT calculations.*

Using a simple Arrhenius law, the desorption rate from a peculiar kink site can be estimated as $k = \alpha \cdot \exp(-\beta E_d)$, where $\alpha$ is the Arrhenius prefactor, $\beta$ the temperature dependent Boltzmann factor and $E_d$ the desorption energy from the considered kink site. Assuming the same prefactor for the desorption events considered here, and using the data reported in Table 1, we obtain that, at 650 K, the sublimation from the Te$_{vdWG}$ site is more than 700 times slower than the sublimation from the Bi$_A$ site. Finally, the largest stability of the Bi$_A$ sites within the $(0\bar{1}10)$ kink sites suggests that $(0\bar{1}10)$ surface layers are Bi-rich during desorption, which is in line with the observed preferential sublimation of Te. The calculations therefore confirm that the sublimation on prismatic $\{01\bar{1}0\}$ planes is far more energetically favoured than the sublimation on the basal planes.

**Table 1**. The desorption energies of kink atoms in the $\{0001\}$ and $\{01\bar{1}0\}$ sites.

| Kink Site   | $\{0001\}$ | Te$_{vdWG}$ | $\{01\bar{1}0\}$ | Te$_A$ | Te$_B$ | Te$_C$ | Bi$_A$ | Bi$_B$ |
|-------------|------------|-------------|------------------|--------|--------|--------|--------|--------|
| Energy (eV) |            | 4.43        |                  | 3.68   | 3.82   | 2.79   | 4.06   | 3.72   |

This also indicates that the reverse process of crystal growth on the $\{01\bar{1}0\}$ type surfaces is easier and faster than on other crystallographic facets. This is in agreement with the plate-like morphology and the $\{01\bar{1}0\}$ planes faceted Bi$_2$Te$_3$ and Bi$_2$Se$_3$ NC produced by both colloidal solution-based and by solid state-based processes. The sublimation mechanism observed here in fact closely mimics, in reverse, the growth mechanism of NCs grown in a solid state system.[12,62]

The sublimation mechanisms of other types of colloidal NCs are more complex due to interplays between a number of different factors affecting their surface properties and each case has to be considered individually. A recent study on the sublimation of the CdSe nanorods for example, showed that basal planes were the ones preferentially sublimating.[31] Although the packing density of the prismatic $\{01\bar{1}0\}$ planes in the CdSe structure is smaller than that of the $\{0001\}$ planes, as in the $Bi_2Te_3$ and $Bi_2Se_3$ NCs, their surface free energy is reduced by the preferential passivation by the surfactant molecules,[63] and the unstable polar $(0001)$ or $(000\bar{1})$ facets were the most prone to sublimation under the pressure and temperature conditions applied.[31] Reduced surface diffusion of atoms due to NC surface stabilization by graphitic shell originating from the surfactant,[30] the surface-stabilizing contact of NCs with the substrate,[34] and the polarity of the structure,[29] have all been shown to affect the sublimation of anisotropic colloidal NCs.

To summarize, the thermal stability of the 2D $Bi_2Te_3$ and $Bi_2Se_3$ platelet-like NCs produced by colloidal synthesis and by LPE was studied during annealing by the *in situ* TEM. Both $Bi_2Te_3$ and $Bi_2Se_3$ undergo sublimation under the vacuum conditions in TEM at temperatures between about 280°C and 500°C, depending on their thickness. The sublimation of the $Bi_2Te_3$ and $Bi_2Se_3$ NCs is also highly anisotropic and involves predominantly the removal of the individual atoms from the prismatic $\{01\bar{1}0\}$ type planes, mimicking in reverse the growth of similar 2D NCs in solid state. This sublimation mechanism is independent of the presence of surfactant (PVP) on the NC surfaces, the method of their preparation and their morphology. The DFT calculations confirm that the sublimation of the $\{01\bar{1}0\}$ planes is energetically less costly and proceeds at least about 700 times faster than the same process on the basal $\{0001\}$ planes. The observed phenomenon is expected to extend to other $Me_2X_3$ (Me-metal; X-chalcogen) NCs adopting the same structure as $Bi_2Te_3$ and $Bi_2Se_3$. The preferential Te and Se sublimation was also observed and these processes may be taking place ahead of, and likely below temperatures at which notable structural disintegration of the NCs begins.

**Author Information**

Corresponding Authors

*E-mail: joka.buha@iit.it; liberato.manna@iit.it

**Notes**

The authors declare no competing financial interests.

**Supporting information.** This material includes: additional details on colloidal and LPE syntheses, surfactant removal procedure, DFT calculations, XRD, Raman, optical absorption and TEM characterizations; additional TEM images following the thermal evolution of the surfactant-free $Bi_2Te_3$ NCs, XRD characterization, as well as Raman and optical absorption spectra of the representative samples. Also included are Video S1 showing the sublimation of a $Bi_2Te_3$ NC with an atomic resolution during the *in situ* annealing in the TEM, and Video S2 showing the sublimation of a LPE $Bi_2Te_3$ flake. Both videos are played at eight times the actual speed. The supporting Information is available free of charge on the ACS Publications website.

ACKNOWLEDGMENTS

This work was supported by the European Union's Seventh Framework Programme FP7/2007-2013 under the ERC research grant TRANS-NANO (contract number 614897) and Graphene Flagship (contract number CNECT-ICT-604391). The IIT platform CompuNet and the Italian Supercomputer Center CINECA (Bologna, Italy) are acknowledged for the computational resources. The authors would like to thank Prof. Andrea Cavalli for discussion on the computational aspects of this study.

REFERENCES

1. Goldsmid, H.J.; Douglas, R.W. *Brit. J. Appl. Phys.* **1954,** 5, 386-390.

2. Goldsmid, H.J.; *Materials* **2014**, 7, 2577-2592.

3. Zhang, H.; Liu, C-X.; Qi, X-L.; Fang, Z.; Zhang, S-C. *Nature Physics* **2009**, 5, 439-442.

4. Hasan, M.Z.; Kane, C.L. *Rev. Mod. Phys.* **2010**, 82, 3045-3067.

5. Cha, J.J.; Koski, K.J.; Cui, Y. *Phys. Status Solidi RRL*, **2012**, 1-11.

6. Hicks, L.D.; Dresseslhaus, M.S. *Phys. Rev. B* **1993**, 47,12727.

7. A. Balandin, K.L. Wang, *Phys. Rev. B* **1998,** 58, 1544-1549.

8. Dresselhaus, M.S.; Chen, G.; Tang, M.Y.; Yang, R.; Lee, H.; Wang, D.; Ren, Z.; Fleurial, J-P.; Gogna, P. *Adv. Mater.* **2007**, 19, 1043-1053.

9. Feutelais, Y.; Legendre, B.; Rodier, N.; Agafonov, V. *Mater. Res. Bull.* **1993**, *28*, 591–596.

10. Nakajima, S. *J. Phys. Chem. Sol.* **1963**, 24, 479-485.

11. Fang, L.; Jia, Y.; Miller, D.J.; Latimer, M.L.; Xiao, Z.L.; Welp, U.; Crabtree, G.W.; Kwok, W.-K. *Nano Lett.* **2012**, 12, 6164-6169.

12. Kong, D.; Wang, W.; Cha, J.J.; Meister, S.; Peng, H.; Liu, Z.; Cui, Y.; *Nano Lett.* **2010**. 10, 2246-2250.

13. Jacobs-Gedrim, R.B.; Durcan, C.A.; Jain, N.; Yu, B. *Appl. Phys. Lett.* **2012**, 101, 143103.

14. Zhang, G.; Qin, H.; Teng, J.; Guo, J.; Guo, Q.; Dai, X.; Fang, Z.; Wu, K. *Appl. Phys. Lett.* **2009**, 95, 053114.

15. Teweldebrhan, D.; Goyal, V.; Balandin, A.A. *Nano Lett.* **2010**, 10, 1209-1218.

16. Ding, Z.F.; Bux, S.K.; King, D.J.; Chang, F.L.; Chen, T-H.; Huang, S-C.; Kaner, R.B. *J. Mater. Chem.* **2009**, 19, 2588-2592.

17. Coleman, J.N.; Lotya, M.;O'Neill, A.;Bergin, S.D.; King, P.J.; Khan, U.; Young, K.; Gaucher, A.; De, S.; Smith, R.J.; Shverts, I.V.; Arora, S.K.; Stanton, G.; Kim, H-Y.; Lee, K.; Kim, G.T.; Duesberg, G.S.; Hallam, T.; Boland, J.J.; Wang, J.J.; Donegan, J.F.; Grunlan, J.C.; Moriarty, G.; Shmeliov, A.; Nicolls, R.J.; Perkins, J.M.; Grieveson, E.M.; Theuwissen, K.; McComb, D.W.; Nellist, P.D.; Nicolosi, V. *Science* **2011**, 331, 568-571.

18. Sun, L.; Lin, Z.; Peng, J.; Weng, J.; Huang, Y.; Luo, Z. *Sci. Reports* **2014**, 4, 4794.

19. Min, Y.; Moon, G.D.; Kim, B.S.; Lim, B.; Kim, J-S.; Kang, C.Y.; Jeong, U.*; J. Am. Chem. Soc.* **2012**, 134, 2872-2875.

20. Min, Y.; Roh, J.W.; Yang, H.; Park, M.; Kim, S.I.; Hwang, S.; Lee, S.M.; Lee, K.H.; Jeong, U. *Adv. Mater.* **2013**, 25, 1425-1429.

21. Kong, D.; Koski, K.J.; Cha, J.J.; Hong, S.S.; Cui, Y. *Nano Lett.* **2013**, 13, 632-636.

22. Xu, H.; Chen, G.; Jin, R.; Chen, D.; Wang, Y.; Pei, J.; Zhang, Y.; Yan, C.; Qiu, Z. *Cryst. Eng. Comm.* **2014**, 16, 3965-3970.

23. Lin, Z.; Chen, Y.; Yin, A.; He, Q.; Huang, X.; Xu, Y.; Liu, Y.; Zhong, X.; Huang, Y.; Duan X. *Nano Lett.* **2014**, 14, 6547-6553.

24. Scheele, M.; Oeschler, N.; Meier, K.; Kornowski A.; Klinke, C.; Weller H. *Adv. Func. Mater.* **2009**, 19, 3476-3483.

25. Min, Y.; Moon, G.D.; Kim, C-E.; Lee, J-H.; Yang, H.; Soon, A.; Jeong, U.; *J. Mater. Chem. C* **2014**, 2, 6222-6248.

26. Scanlon, D.O.; King, P.D.C.; Singh, R.P.; de la Torre, A.; McKeown Walker, S.; Balakrishnan, G.; Catlow, C.R.A. *Adv. Mater.* **2012**, 24, 2154-2158.

27. Talapin, D.V.; Lee, J-S.; Kovalenko, M.V.; Shevchenko, E. V. *Chem. Rev.* **2010**, 110, 389-458.

28. Carey, G.H.; Abdelhady, A.L.; Ning, Z.; Thon, S.M.; Bakr, O.M.; Sargent, E.H. *Chem. Rev.* **2015**, 115, 12732-12763.

29. van Huis, M.A.; Young, N.P.; Pandraud, G.; Creemer, J.F.; Vamaekelbergh, D.; Kirkland, A.I.; Zandbergen, H.W. *Adv Mater.* **2009**, 21, 4992-4995.

30. Khalavka, Y.; Olm, C.; Sun, L.; Banhart, F.; Sönnichsen, C. *J. Phys. Chem. C* **2007**, 111, 12886-12889.

31. Hellebusch, D.J.; Manthiram, K.; Baberwyck, B.J.; Alivisatos, A.P. *J. Phys. Chem. Lett.* **2015**, 6, 605-611.

32. Yim, J.W.L.; Xiang, B.; Wu, J. *J. Am. Chem. Soc.* **2009**, 131, 14526-14530.

33. Link, S.; Wang, Z.L.; El-Sayed, M.A. *J. Phys. Chem. B* **2000**, 104, 7867-7870.

34. Goris, B.; Van Hius, M.A.; Bals, S.; Zandbergen, H.W.; Manna, L.; Van Tendeloo, G. *Small* **2012**, 8, 937-942.

35. Liu, X.; Wood, J.D.; Chen, K-S.; Cho, EK.; Hersam, M.C. *J. Phys. Chem. Lett.* **2015**, 6, 773-778.

36. Huang, J.Y.; Ding, F.; Yakobson, B.I.; Lu, P.; Qi, L.; Li, J. *PANS* **2009**, 106, 10103-10108.

37. Liu, Z.; Suenaga, K.; Harris, P.J.F.; Iijima, S. *Phys. Rev. Lett.* **2009**, 102, 015501.

38. Koczkur, K.M.; Maurdikoudis, S.; Polavarapu, L.; Skrabalak, S.E. *Dalton Trans.* **2015**, 44, 17883-17905.

39. Du, Y.K.; Yang, P.; Mou, Z.G.; Hua. N.P.; Jiang, L. *J. Appl. Polymer Sci.* **2005**, 99, 23-26.

40. Ding, Y.; Fan, F.; Tian, Z.; Wang, Z.L. *Small* **2009**, 5, 2812-2815.

41. Zhang, G.; Kirk, B.; Jauregui, L.A.; Yang, H.; Xu, X.; Chen, Y.P.; Wu, Y. *Nano Lett.* **2012**, 12, 56-60.

42. Zhou, K.; Mao, N.; Wang, H.; Peng, Y.; Zhang, H. *Angew. Chem. Intl. Ed.* **2011**, 50**,** 10839-10842.

43. Halim, U.; Zheng, C. R.; Chen, Y.; Jiang, S.; Cheng, R.; Huang, Y.; Duan, X. *Nature Commun.* **2013**, 4, 3213.

44. Hohenberg, P.; Kohn, W. *Physical Review* **1964,** 136, 3B, 864-871.

45. Kohn, W.; Sham, L. J. *Physical Review* **1965,** 140, 4A, 1133-1138.

46. Honig, R.E. *RCA Review*, **1957**, June, 195-204.

47. Brebrick, R.F.; Smith, T.J. *J. Electrochem. Soc.* **1971**, 118, 991-996.

48. Lee, J.; Berger, A.; Cagnon, L.; Gösele, U.; Nielsch, K.; Lee, J. *Phys. Chem. Chem. Phys.* **2010**, 12, 15247-15250.

49. Brostow, W.; Datashvili, T.; Hagg Lobland, H.E.; Hilbig, T.; Su, L.; Vinado, C.; White, J.; *J. Mater. Res.* **2012**, 27, 2930-2936.

50. Hassel, O. *Z. Phys.* **1924**, 24, 317-337.

51. Stachurski, Z.H. *Materials* **2011**, 4, 1564-1598.

52. He, X.; Zhang, H.; Lin, W.; Wei, R.; Qui, J.; Zhang, M.; Hu, B. *Sci. Rep.* **2015**, 5, 15868.

53. Tagaki, M. *J. Phys. Soc. Jpn.* **1956**, 9, 395-363.

54. Hanszen, K.-J. *Z. Phys.* **1960**, 157, 523-553.

55. Asoro, M.A.; Kovar, D.; Ferreira, P.J. *ACS Nano* **2013**, 7, 7844-7852.

56. Furukawa, Y.; Nada, H. *J. Phys. Chem. B* **1997**, 101, 6167-6170.

57. Sankarasubramanian, R.; Kumar, K. *Computational Mater. Sci.* **2010**, 49, 386-391.

58. Trittibach, R.; Grütter, Ch.; Bilgram, J. H. *Phys. Rev. B* **1994**, 50, 2529-2536.

59. Lindemann, C. L. *Phys. Z.* **1911**, 12, 1197-1199.

60. Jayanthy, C.S.; Tosatti, E.; Pietronero, L. *Phys. Rev. B* **1985**, 31, 3456-3459.

61. Chatterjee, B. *Nature* **1978**, 275, 203.

62. Alegria, L.D.; Schroer, M.D.; Chatterjee, A.; Poirier, G.R.; Pretko, M.; Patel, S.K.; Petta, J.R. *Nano Lett.* **2012**, 12, 4711-4714.

63. Manna, L.; Wang, L. W.; Cingolani, R.; Alivisatos, A.P. *J. Phys. Chem. B* **2005**, 109, 6183-6192.

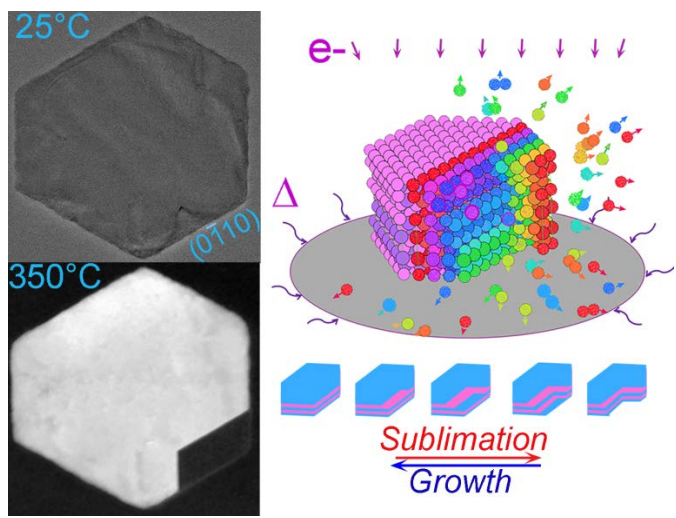

TOC graphic